\documentclass[twocolumn,prl,showpacs,amsmath,amssymb]{revtex4}
\usepackage{graphicx,color}
\usepackage{hyperref}
\begin{document}
\title{Adiabatic Pumping in Interacting Systems}
\author{Eran Sela and Yuval Oreg \date{\today}}
\affiliation{Department of Condensed Matter Physics, Weizmann
Institute of Science, Rehovot, 76100, ISRAEL }

\begin{abstract}
A dc current can be pumped through an interacting system by
periodically varying two independent parameters such as magnetic
field and a gate potential. We present a formula for the adiabatic
pumping current in general interacting systems, in terms of
instantaneous properties of the system, and find the limits for
its applicability. This formula generalizes the scattering
approach for noninteracting pumps. We study the pumped spin in a
system that exhibits the two-channel Kondo effect as an
application of the adiabatic pumping formula. We find that a
quantized spin of $\hbar$ is transferred between the two channels
as the temperature approaches zero, and discuss the non-Fermi
liquid features of this system at finite temperatures.
\end{abstract}
\pacs{72.25.-b, 73.23.-b}

\maketitle

{\it Introduction and Conclusions}.---The scattering approach of
Brouwer~\cite{Brouwer98} for pumping through a finite, possibly
disordered region of noninteracting electrons, which followed a
work by B\"{u}ttiker, Pr\^{e}tre and Thomas~\cite{Buttiker94}
(BPT), tremendously enhanced the understanding of time dependent
transport in mesoscopic systems. The derivation of the Brouwer
formula~\cite{Brouwer98} and its applications are well
established~\cite{Avron02,Entin02}, and recent studies considered
special cases of interactions in quantum
dots~\cite{Aleiner98a,Fazio05short} and in Luttinger
liquids~\cite{Sharmab03,Sharmaplus03}. However, an apt formulation
of the pumped current in the general case when interactions
between the electrons are involved (beyond the Hartree level) has
been lacking.

In this Letter we develop a generic formula [Eq.~(\ref{eq:kub})]
which expresses the pumped current through a region of interacting
electrons in the adiabatic limit. The pumped current is expressed
in terms of an instantaneous linear response function, calculated
at every step along the pumping trajectory. When the motion along
the trajectory is sufficiently slow, the current can be found by
integrating the contributions of each step [see
Eq.(\ref{eq:gre})]. We also rewrite Eq.~(\ref{eq:kub}) for the
case of a quantum dot connected to noninteracting leads in terms
of dot properties [Eq.~(\ref{eq:eml})], from which the
noninteracting S-matrix formula~\cite{Buttiker94} follows as a
special case.

Finally, we consider as an example spin pumping in the vicinity of
the two-channel Kondo (2CK) fixed point which is perturbed by two
pumping parameters, magnetic field and channel anisotropy. Due to
the non Fermi liquid (NFL) nature of the 2CK fixed point, the
standard scattering approach can not be applied straightforwardly.
Using Eq.~(\ref{eq:kub}) we calculate the spin pumped from one
channel into the other, as function of the parametric trajectory
and temperature $T$. We find that as $T \rightarrow 0$ the pumped
spin is quantized in units of $\hbar$ per period for trajectories
which surround the NFL fixed point [paragraph following
Eq.~(\ref{eq:res})], while at finite temperatures this
quantization is not accurate [Eq.~(\ref{eq:spo})]. At finite
temperature, when the trajectory is sufficiently close to the
fixed point, the temperature dependence of the pumped spin
reflects the NFL physics of the 2CK [Eq.~(\ref{eq:sqr})].

{\it Generic adiabatic pumping formula}.---We analyze a mesoscopic
conductor which is described by an Hamiltonian $H_{X_1,X_2,\dots}$
that depends at least on two external parameters $X_1(t)$ and
$X_2(t)$ which are varied periodically and slowly in time. These
parameters can be for example two metallic gates and/or an
external magnetic field. By assumption the conductor may be
divided into left ($L$) and right ($R$) contacts whose Hamiltonian
does not depend on the parameters, and a central region whose
Hamiltonian does depend on the parameters. Interactions may take
place everywhere in the conductor.

In a pumping cycle the set of the parameters $\vec
X(t)=(X_1(t),X_2(t),\dots)$ is varied periodically in time and
defines a closed trajectory, $\mathcal{L}$, in the parameter space.
The change in the parameters may produce a current $J_j$ in
contact-$j \ (j=L,R)$ \cite{Remark05b}. To calculate $J_j$ at a
point $\vec X_0$ along the trajectory due to a small and slow
change, $\delta X$, of one of the parameters, we assume that $\delta
X= \delta X_{\Omega} e^ {i \Omega \tau}$.

The current $J_j( \Omega )$ is then found by a linear response
perturbation theory with respect to the infinitesimal perturbation
$H^\prime=\frac{\partial H }{\partial X} \delta X_\Omega e^{i
\Omega \tau}$. Including the possibility that the current
operator, $\hat{J}_j$, may depend explicitly on $X$, we obtain the
total response to linear order in $\delta X_\Omega$:
\begin{eqnarray}
\label{eq:jom} \frac{J_j( \Omega )}{ \delta X_\Omega}=\left(\!
\left\langle \frac{\partial \hat{J}_j}{\partial X} \right\rangle -
\frac{i}{\hbar} \int_{-\infty}^0 \!\!\!\!\!\!\!\! d \tau e^{i
\Omega^- \tau} \left\langle \! \left[\hat{J}_j(0), \frac{\partial
H }{\partial X} (\tau) \right] \! \right\rangle \! \right)\!,
\end{eqnarray}
with $\Omega^- =\Omega-i0^+$. Since Eq.~(\ref{eq:jom}) for the
current is a first order expansion in the harmonic perturbation,
the time evolution should be understood as $\hat{O} (\tau)= e^{i
H_{\vec X_0} \tau } \hat{O} e^{-i H_{\vec X_0} \tau }$ and the
quantum averages are performed with ($\Omega$ independent)
instantaneous eigenstates of $H_{\vec X_0}$~\cite{Remark05a}.

The charge $\delta  Q(j,\Omega)= J_j(\Omega)/(i \Omega)$ entering
the central region through contact $j$ is $\delta Q(j,\Omega)= e
\frac{d n(j)}{d X} \delta X_{\Omega} $, where $\frac{d n(j)}{d
X}$, the \emph{emissivity} into contact $j$, is given by:
%\label{eq:emc}
\begin{equation}
\frac{d n(j)}{d X}=\lim_{\Omega \rightarrow 0}\frac{J_j( \Omega
)}{i e \Omega \delta X_\Omega} = \frac{1}{i e \delta X_\Omega}
\left. \frac{d}{d \Omega} J_j( \Omega ) \right|_{\Omega=0}.
\nonumber
\end{equation}
The last equality follows since in the static limit, $\Omega=0$,
no current flows through the contacts. This
yields~\cite{Remark05a,Remark05f}:
\begin{equation}
\label{eq:kub} \frac{d n(j)}{d X}=\frac{1}{ \hbar e }\lim_{\Omega
\rightarrow 0} \frac{d}{ d \Omega } \int_{-\infty}^0
\!\!\!\!\!\!\!d \tau e^{i \Omega^- \tau} \!\! \left\langle \!
\left[\frac{\partial H_{\vec X_0} }{\partial X} (\tau),
\hat{J}_j(0) \right]\! \right\rangle_{\!\!\! \displaystyle{.}}
\end{equation}
Notice that while here we take the limit $\Omega \rightarrow 0$,
in practice the emissivity does not depend on $\Omega$ for $\Omega
<1/\tau_r$, where $\tau_r$ is a characteristic relaxation time.
The dependence of the  emissivity on time is only through the
location of $\vec X_0$ on the trajectory~$\mathcal{L}$.

Using Eq.~(\ref{eq:kub}) for the emissivity related to each
parameter, the charge pumped per period, corresponding to a
trajectory $\mathcal{L}$ in the parameter space, is given
by~\cite{Brouwer98}
\begin{equation}
\label{eq:gre} Q(j)= \int_\mathcal{L} d \vec{X} \cdot \vec{A}(j)=
\int_\mathcal{S} dX_1 dX_2 B(j),
\end{equation}
where $\mathcal{S}$ is the area bounded by $\mathcal{L}$ and the
effective ``vector potential" and ``magnetic field" are
$\vec{A}(j)=e \frac{d n(j)}{d \vec{X}}$ and $B(j)=(\vec{\nabla}
\times \vec{A}(j))_3=\frac{\partial}{\partial X_1} \frac{d n(j)}{d
X_2} -\frac{\partial}{\partial X_2} \frac{d n(j)}{d X_1}$
respectively.

To determine the validity regime of Eqs.~(\ref{eq:kub}) and
(\ref{eq:gre}) we divide the trajectory $\mathcal{L}$ into
elements of length $\delta X $. The length, $\delta X$, should be
smaller than both the radius of curvature along the trajectory,
$r_c= \left|\dot{\vec X}\right|^3\left/\left|\dot{\vec X} \times
\ddot{\vec X}\right|\right.$, and $r_2=\left|\frac{dn}{d X_i}
\dot{\hat{X}}_i \left/ \frac{d^2n}{ dX_k dX_j}
\dot{\hat{X}}_k\hat{\dot{X}}_j\right. \right|$, the length at
which the second order term in $\delta X$ is comparable to the
first order term. Here $\dot{ \hat {X}}_i= \dot {X}_i \left/
\left|\dot {\vec X}\right|\right.$, and the Einstein summation
convention is understood. In addition, $\delta X = \left|\dot{\vec
X} \right| \delta t$, and $\delta t$ must be longer than $\tau_r$,
so that after each step the system relaxes to a new equilibrium
position determined by the Hamiltonian with the new parameters
$\vec X_0+d \vec{X}/d t~\delta t$. Combining both requirements we
find that for every $\vec{X} \in \mathcal{L}$, the length $\delta
X$ has to satisfy: $\left|\frac{d \vec X}{d t}\right| \tau_r \ll
\left|\frac{d \vec X}{d t}\right| \delta t = \delta X \ll \min
\left\{r_2, r_c\right\}$. Thus, formulas~(\ref{eq:kub}) and
(\ref{eq:gre}) are valid for
\begin{equation}
\forall_{\vec{X} \in \mathcal{L}} \left|\frac{d \vec X}{d
t}\right| \ll \frac{1}{\tau_r} \min \left\{r_2, r_c\right\}.
\nonumber
\end{equation}
For a circular trajectory: $\vec X(t) = r_0 \left(\cos \Omega_0 t,
\sin \Omega_0 t \right)$, one easily finds that $r_c=r_0,\left| d
\vec X\left/d t \right. \right|  = \Omega_0 r_0$ and the condition
becomes: $ \Omega_0 \ll \frac{1}{\tau_r} \min \left\{r_2/r_0,
1\right\}$.

{\it Adiabatic pumping formula for quantum dots}.---Next we
consider a quantum dot coupled to noninteracting contacts,
described by a parametric dependent version of the Hamiltonian
discussed in Ref.~\cite{Meir92}, $H_{\vec X}=\sum_{k,\alpha \in
L,R} \epsilon_{k \alpha} c^\dagger_{k \alpha}c_{k \alpha}+H^{\rm
int}_{\vec X}(\{ d_n^\dagger , d_n \}) +\sum_{k,\alpha \in L,R}
V_{k \alpha,n}( \vec X ) c^\dagger_{k \alpha} d_n + h.c$. Here
$c^\dagger_{k \alpha}$ creates an electron with momentum $k$ in
channel $\alpha$ belonging to contact $j$, and $\{ d^\dagger_{n}
\}$ form a complete, orthonormal set of single-electron creation
operators in the dot. In this case the current operator is $\hat
J_{j}= \frac{i e}{\hbar} \sum_{k, \alpha \in j} V_{k \alpha,n}(
\vec X )c^\dagger_{k \alpha} d_n+ h.c.$. A straight forward
calculation shows that using Eq.~(\ref{eq:kub}) the emissivity can
be written as ($\hbar=1$):
\begin{eqnarray}
\label{eq:em2} \frac{d n(j)}{d X}= \frac{d}{
 d\Omega } \sum_{k,\alpha \in j}
\int_{-\infty}^{\infty} \frac{d \omega}{2 \pi i} V_{k \alpha,n}
\delta_X G^{ <}_{n,k \alpha} (\omega,\Omega)+c.c.
\end{eqnarray}
where $u=t_1-t_1'$, $T=\frac{t_1+t_1'}{2}$,
\begin{eqnarray}
\delta_X G^{\eta \eta'}_{a,b}(\omega,\Omega)&=&e^{- i \Omega T}
\int d u e^{i \omega u}\delta_X G_{a,b}^{\eta
\eta'}(t_1,t_1',\Omega),
\nonumber \\
\delta_X G_{a, b}^{ \eta \eta'}(t_1,t_1',\Omega)&\equiv&
\frac{d}{d \delta X_\Omega} G_{a, b}^{\eta
\eta'}(t_1,t_1',\Omega)|_{\delta
X_\Omega=0},~ \text{and} \nonumber \\
G^{\eta \eta'}_{a, b}(t_1,t_1',\Omega)&=&-i \left\langle {\cal
T}_C \left[ e^{-i \int_C d \tau H'(\tau)} a(t_1) b^\dagger(t_1')
\right]\right\rangle. \nonumber
\end{eqnarray}
Here $G^{\eta \eta'}_{a, b}(t_1,t_1',\Omega)$ is the Keldysh Green
function related to the operators $a,b=d_n, c_{k \alpha}$
\cite{Remark05c}. We note that the dependance of the Green
function on $\delta X_\Omega$ is through $H'(\tau)$ defined above
Eq.~(\ref{eq:jom}).

Since the contacts are noninteracting, the summation over $k$ in
Eq.~(\ref{eq:em2}) can be carried out \cite{Meir92}, and the
emissivity can be written in terms of $\delta_X G^{\eta
\eta'}_{n,m}$ only. Defining a vertex function
%\label{eq:dy1}
\begin{equation}
 \delta_X
\hat{\mathbf{G}}(\epsilon,\Omega)=\hat{\mathbf{G}}(\epsilon-\frac{\Omega}{2})
\hat{\mathbf{\Lambda}}_X (\epsilon,\Omega)
\hat{\mathbf{G}}(\epsilon+\frac{\Omega}{2}), \nonumber
\end{equation}
where we used matrix notation both for
the dot indices (bold letters) and for the Keldysh indices (hat, $\hat{~}$), with $\hat{\mathbf{G}}=\left(%
\begin{array}{cc}
  \mathbf{G}^r & \mathbf{G}^< \\
  0 & \mathbf{G}^a \\
\end{array}%
\right)$, we find~\cite{Fazio05short,Remark05d}

\widetext
\begin{equation}
\label{eq:eml}  \frac{d n(j)}{d X} =-\frac{\partial n_j}{
\partial X}+  \int
\frac{d \epsilon} {2 \pi} \text{tr} \left\{ \!\frac{d f}{d
\epsilon} \text{Re} \bigl[
 \frac{\partial (\mathbf{\Gamma}^j
\mathbf{G}^r)}{\partial X} {{\mathbf{G}^r}}^{-1} \mathbf{G}^a
+\bar
\partial_X \mathbf{\Gamma}^j \mathbf{G}^r \bigr]  \ +\mathbf{G}^a
\mathbf{\Gamma}^j \mathbf{G}^r \frac{d}{d \Omega}[
(\mathbf{\Lambda}_X^r-\mathbf{\Lambda}_X^a)f+\mathbf{\Lambda}_X^{<}]_{\Omega=0}
\right\}.
\end{equation}
\endwidetext
The arguments of $f(\epsilon)$, $\mathbf{G}(\epsilon)$,
$\mathbf{\Gamma}(\epsilon)$, and
$\mathbf{\Lambda}_X(\epsilon,\Omega)$ were suppressed.
$\protect{n_j = \sum_{\alpha \in j, k} \int d \omega f (\omega)
\big[-\frac{1}{\pi} \text{Im} G_{k \alpha, k \alpha}(\omega)\big]
}$ is the equilibrium occupancy of contact $j$ calculated in the
presence of the dot, $\Gamma^j_{n,m}(\epsilon)=2 \pi \sum_{\alpha
\in j}  V_{n, \alpha}(\epsilon)\rho_\alpha(\epsilon) V_{\alpha,
m}(\epsilon) $, $\rho_\alpha(\epsilon)$ is the bare density of
states in channel $\alpha$ and $V_{\alpha,n}(\epsilon)=V_{k
\alpha, n}$ for $\epsilon=\epsilon_{k \alpha}$. The matrix
$\protect{ \bar
\partial_X \Gamma^j_{n,m} \equiv \frac{\overrightarrow{\partial}}{\partial X}
\Gamma^j_{n,m} -\Gamma^j_{n,m}
\frac{\overleftarrow{\partial}}{\partial X}}$ is antihermitian.

In the noninteracting case one can show that expression
(\ref{eq:eml}) without the first term, $-\frac{\partial
n_j}{\partial X}$, reduces to BPTs' result. The BPTs' emissivity
contains the explicit derivative contribution $+\frac{\partial
n_j}{\partial X}$, as they calculate the current deep inside the
reservoir, while in this Letter the current defined above
Eq.~(\ref{eq:em2}) is calculated at the entrance to the dot. An
explicit derivative does not influence the pumping charge per
period, and formally corresponds to a gauge transformation in the
vector potential defined after Eq.~(\ref{eq:gre}). Notice that for
an infinite flat band and energy independent tunneling couplings
$\frac{\partial n_j}{\partial X} =0$ even when interactions in the
dot are included~\cite{Hewson93d}.

{\it Pumping in the two channel Kondo effect}.---To demonstrate
the new features of Eq.~(\ref{eq:kub}) which includes
interactions, we study a specific example of pumping in a 2CK
system at the exactly solvable Emery-Kivelson (EK)
line~\cite{Emery92}. The peculiarity of a \emph{symmetric} 2CK
problem is in its NFL behavior at low
temperatures~\cite{Nozieres80}. In the presence of external
magnetic field, $B$, the Hamiltonian of the 2CK model is
\begin{equation}
 H^{\rm 2CK}=\sum_{k \sigma j } \epsilon_k c_{k
\sigma j}^\dagger c_{k \sigma j}+  \sum_{j,\lambda} \mathcal{J}_{j
\lambda} S_\lambda \cdot s_j^\lambda + g \mu_B H S_z. \nonumber
\end{equation}
The index $j=1,2$ represents two channels, $\vec S$ is the
impurity spin-$1/2$-operator, $s_j^\lambda$ is the spin density in
channel $j$ in direction $\lambda=x,y,z$ near the localized spin.

In the present context we consider \emph{spin} pumping from
channel 1 into channel 2 by calculating the spin-flavor (sf)
emissivity $\frac{d s_{1 \rightarrow 2}}{d X} \equiv
\frac{\hbar}{2} \bigl( \frac{d s_z(1)}{d X}-\frac{d s_z(2)}{d X}
\bigr)$, using Eq.~(\ref{eq:kub}) with $\hat{J}_j$ replaced by
$\hat{J}_{\rm sf} = -e \frac{d n_{\rm sf}}{dt} =\frac{i e}{\hbar}
[n_{\rm sf},H^{\rm 2CK}]$, where $n_{\rm sf} = s_z(1)-s_z(2)$ is
the spin difference between the channels. At the EK line
$\mathcal{J}_{1 z} =\mathcal{J}_{2 z} =2 \pi \hbar v_F$ the spin
charge and flavor sectors commute with the spin-flavor (sf) sector
that determines the evolution of $\hat{J}_{\rm sf}$.

In the presence of  channel anisotropy in the spin flip processes,
$\mathcal{J}_{1 \perp} \ne \mathcal{J}_{2 \perp}$, the Hamiltonian
of the SF sector
in a Nambu notation: $\Psi^\dagger_i=\left(%
  \psi^\dagger_i,
  \psi_i \right)$, $i=d,\rm{sf}$ takes the form of a Majorana resonance level (MRL)
model
\begin{eqnarray}
 H^{\rm MRL}_{h,\Delta}=  \frac{1}{2} ( i v_F
\int_{-\infty}^\infty dx \Psi_{\rm sf}^\dagger(x) \hat{\tau}_z
\frac{ \partial \Psi_{\rm
sf}(x)}{\partial x} \nonumber \\
 + \frac{\Gamma}{2} h
\Psi_d^\dagger \hat{\tau}_z \Psi_d +\Psi_{\rm sf}^\dagger(0)
\hat{V}^\dagger \Psi_d + \Psi_d^\dagger \hat{V} \Psi_{\rm sf}(0)),
\nonumber
\end{eqnarray}
where  $\hat V = \sqrt{\Gamma/(2 \pi \rho)} \left(\cos (\theta/2)
\tau_z+i \sin (\theta/2) \tau_y \right)$, $\Gamma \equiv
\Gamma_1+\Gamma_2$, $\Gamma_{1(2)}=\rho \mathcal{J}_{1(2) \perp}^2
/4 a $, $a$ can be considered as the lattice spacing, $h= 2 g
\mu_B H/\Gamma$, $\rho=(2 \pi \hbar v_F)^{-1}$ is the density of
states of the chiral fermion field $\psi_{\rm sf}$, and $\vec
\tau$ are the pauli matrices. The operator that describes
transition of spin between the channels is $n_{\rm sf} =
\int_{-\infty}^\infty dx \psi_{\rm sf}^\dagger(x)  \psi_{\rm
sf}(x)$~\cite{Delft98} and $\psi_d$ is a local fermion operator.

The unique properties of the 2CK system can be seen by taking
$\theta=\pi/2$ and $h=0$ in the MRL model. In order to describe
pumping in the vicinity of this point we choose as pumping
parameters $X_1 =h$ and
$X_2=\Delta=(\Gamma_1-\Gamma_2)/{\Gamma}=\cos \theta$.

Using Eq.~(\ref{eq:kub}) one finds that the spin flavor emissivity
is given by $\vec{A}=\frac{d s_{1 \rightarrow 2}}{d \vec{X}}=-
\int_{-\infty}^{\infty} d \epsilon \frac{df(\epsilon)}{d \epsilon}
\vec{a}(\epsilon)$, where
\begin{equation}
\label{eq:res} \left(a_h,a_\Delta\right) = \frac{\hbar}{2 \pi}
\frac{r^2 (-\Delta,h)-4 \tilde\epsilon^2  \left(\Delta,h\right)- 4
\tilde\epsilon (h,\Delta)}{16 \tilde\epsilon^4+8 \tilde\epsilon^2
(2-r^2) + r^4}.
\end{equation}
Here $\tilde \epsilon = \epsilon/\Gamma$, $r =
\sqrt{h^2+\Delta^2}$, and $\frac{d }{d \vec{X}}\equiv (\frac{d }{d
h},\frac{d }{d \Delta})$.
\begin{figure}[h]
\begin{center}
\includegraphics*[width=80mm,height=60.3279mm]{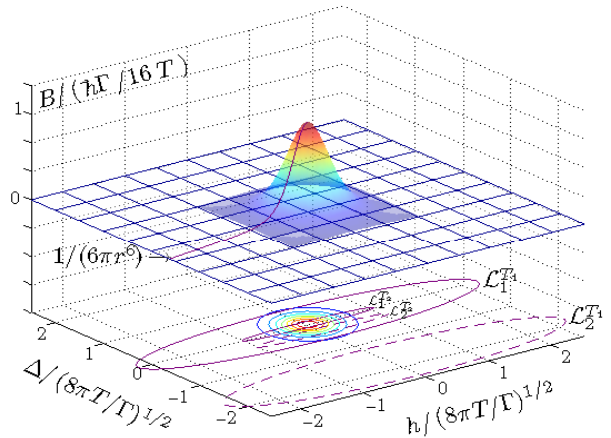}
\caption{The effective magnetic field $B$ in the $(h,\Delta)$
parameter space. For $T \ll \Gamma$ the effective magnetic field
has a peak of weight $\hbar$ at $(h,\Delta)=(0,0)$ and the shape
of the peak is temperature independent if plotted against the
scaled parameters $(h /\sqrt{8 \pi T / \Gamma},\Delta/\sqrt{8 \pi
T / \Gamma})$.
 Each of the pumping trajectories $\mathcal{L}_{1}:$ $
\left({h}/{h_0}\right)^2+({\Delta}/{\Delta_0})^2=1$ (full line)
and $\mathcal{L}_{2}:$ $ \left({(h+3
h_0)}/{h_0}\right)^2+({\Delta}/{\Delta_0})^2=1$ (dashed line) is
plotted for $T_1 \ll \Gamma \Delta_0^2$ and for $T_2 \gg \Gamma
\Delta_0^2$ in the scaled parameter space, leading to four
different curves $\mathcal{L}_i^{T_j}$, $i,j=1,2$ (all
trajectories are anticlockwise). Denoting the area bounded by
$\mathcal{L}_i^{T_j}$ as $\mathcal{S}_i^{T_j}$, we see that while
$\mathcal{S}_1^{T_1}$ contains the entire area of the peak of $B$,
correspoding to pumping a spin of exactly $\hbar$ from channel 1
to 2, the intersection of the area of the peak with
$\mathcal{S}_2^{T_1}$ is empty. On the other hand, the relative
area of the peak contained in $\mathcal{S}_{1,2}^{T_2}$ is
approximatelly $ \frac{\Delta_0 \sqrt{T / \Gamma}}{T/\Gamma}$,
corresponding to the pumped spin $\sim \hbar \Delta_0   /
\sqrt{T/\Gamma}$. The anomalous power is a manifestation of the
NFL behavior. \label{fg:3}}
\end{center}
\end{figure}

At $T=0$ Eq.~(\ref{eq:res}) gives $\vec{A}=\frac{\hbar}{2
\pi}\frac{(-\Delta, h)}{h^2+\Delta^2}$, (which is $\frac{\hbar}{2
\pi} \frac{\hat{\phi}}{r}$ in polar coordinates) -- the vector
potential corresponding to an effective magnetic field $B= \hbar\
\delta^2 (h,\Delta)$ perpendicular to the plane of the trajectory.
Thus, in a pumping period encircling the NFL point
$(h,\Delta)=(0,0)$ anticlockwise, a total spin of exactly $\hbar$
is transferred from channel 1 to channel~2~\cite{Remark05k}. The
magnetic field $B$ can be obtained analytically for any finite
temperature. We discuss below the different regions of $B$ as a
function of $T$.

For $T/\Gamma \ll 1$, namely for temperatures smaller than the
Kondo scale, the pumping vector potential can be approximated by
$\vec{a}(\epsilon)=\frac{\hbar}{2 \pi} \frac{r^3 \hat{\phi}}{16
\tilde{\epsilon}^2+r^4}$, which after integration over $\epsilon$
gives $\vec{A}= \frac{\hbar r \hat{\phi } \Gamma }{16 \pi^2 T}
\psi_1(\frac{1}{2}+\frac{r^2 \Gamma}{8 \pi T})$, where $\psi_1$ is
the trigamma function. (We have neglected the third term in
Eq.~(\ref{eq:res}) which is curl-less.) The magnetic field $B=
\vec \nabla \times \vec A$ is
\begin{equation}
\label{eq:bps} B(r)= {\hbar \Gamma }/{( 8 \pi  T)}
F_1\left({r}/{\sqrt{8 \pi T/\Gamma}}\right),
\end{equation}
with the function $ F_1(x)=\frac{1}{\pi}\frac{\partial}{\partial
y}(y \psi_1(\frac{1}{2}+y)) |_{y=x^2}$ satisfying $F_1(0)=\pi/2$
and $F_1(x) \xrightarrow[] {x \gg 1 }\frac{1}{6 \pi x^6}$. This
means that the effective magnetic field of strength $ \sim \hbar
\Gamma /(16 T)$ is concentrated in a circle of radius $\sim \sqrt
{T/\Gamma}$, and decays strongly as $\hbar \frac{32 \pi}{3}
\frac{\left(T/\Gamma\right)^2}{r^6}$, as depicted in
Fig.~\ref{fg:3}.

At $T \sim \Gamma$ the peak size approaches unity and ceases to be
circularly symmetric due to the second term in Eq.~(\ref{eq:res}).

For $T \gg \Gamma$ the effective magnetic field becomes
practically independent of $-1 \leq \Delta \leq 1$, and given by
\begin{equation}
\label{eq:hig} B(h,T)=\frac{\hbar \Gamma}{16 T}
F_2\left(h\frac{\Gamma}{T}\right),
\end{equation}
with $F_2(x)=\frac{2}{\pi^2} \text{Re}\left[
\psi_1(\frac{1}{2}-\frac{x}{4 \pi i}) \right]$. Since
$\protect{\int_{-\infty}^{\infty} dx F_2(x) = 8}$, the weight of
the pumping peak is again $\hbar$, however the peak is very wide
$\sim T/\Gamma \gg 1$.

The total pumped spin is obtained by performing the integral $s_{1
\rightarrow 2}^{\mathcal{L}}=\int_{\cal S} d^2r B(r)$, where
${\cal S}$ is the area contained in the trajectory $\mathcal{L}$.
We can easily estimate the temperature dependance of the pumped
spin, using the structure of $B(r)$ described in
Eqs.~(\ref{eq:bps}) and (\ref{eq:hig}).

Consider for example, $\mathcal{L}_1$, an elliptic pumping
trajectory $ \left({h}/{h_0}\right)^2+({\Delta}/{\Delta_0})^2=1$
where $\Delta_0 \ll 1 \ll h_0  $ (see Fig.~\ref{fg:3}). At $T=0$
it encircles the origin and therefore $s_{1 \rightarrow
2}^{\mathcal{L}_1}(T=0)=\hbar$. At low temperatures $\sqrt{T
/\Gamma} \ll \Delta_0 $, taking into account the tail of the peak
of $B$, we obtain
\begin{equation}
\label{eq:spo} s_{1 \rightarrow 2}^{\mathcal{L}_1}(T) =\hbar
\left(1-\left(\frac{T}{T_{0}}\right)^2\right),\; T_{0} =
\frac{\Gamma}{4\pi} \sqrt{\frac{6 h_0^3
\Delta_0^3}{h_0^2+\Delta_0^2}}.
\end{equation}
This is expected from the FL behavior along the trajectory (no
anomalous exponents appear). At higher temperatures when $\Delta_0
\ll \sqrt{T /\Gamma} \ll 1$ we find
\begin{equation}
\label{eq:sqr} s_{1 \rightarrow 2}^{\mathcal{L}_1}(T) = c \ \hbar
\sqrt{{ \Gamma}/{T}} \Delta_0,
\end{equation}
with $c$ of order unity. Here the anomalous exponents ($\sim
T^{-1/2}$) of the NFL point become apparent. In Fig.~\ref{fg:3}
the magnetic field is plotted in terms of scaled parameters $(h
/\sqrt{8 \pi T / \Gamma},\Delta/\sqrt{8 \pi T / \Gamma})$. Using
this scaled parameters, as long as $T \ll \Gamma$, the shape of
the peak is temperature independent, however each trajectory
$\mathcal{L}$ acquires a temperature dependence~$\mathcal{L}^T$.
Let us consider the temperatures $T_1$ and $T_2$ satisfying
$\sqrt{T_1 /\Gamma} \ll \Delta_0 $ and $\Delta_0 \ll \sqrt{T_2
/\Gamma} \ll 1$ respectively. We see that the area bounded by
$\mathcal{L}_1^{T_1}$ contains the entire peak of $B$ while the
area bounded by $\mathcal{L}_1^{T_2}$ contains only a one
dimensional cut of the peak, (whose radius scales as $\Gamma/T$
for the bare parameters) explaining the anomalous behavior for
$T_2$.

For $1 \ll T /\Gamma \ll h_0$ we find $s_{1 \rightarrow
2}^{\mathcal{L}_1}(T) = \hbar \Delta_0$. Finally for very large
temperatures $1 \ll h_0 \ll T/\Gamma$ we have $s_{1 \rightarrow
2}^{\mathcal{L}_1}(T)=\hbar \pi h_0 \Delta_0 \Gamma / (16 T)$.
Such area law [ $O(\pi h_0 \Delta_0)$] is expected in this regime,
since $B$ is practically constant for $h_0,\Delta_0 \ll T /
\Gamma$.

We acknowledge useful discussions with Natan Andrei and Alessandro
Silva. This research was supported by the following grants: ISF
845/04, Minerva, DIP C71 and BSF 2004278.

\vspace{-0.5cm}
%\bibliography{library,lib1}

\begin{thebibliography}{20}
\bibitem[{Rem({\natexlab{a}})}]{Remark05b}
\bibinfo{note}{Here $J_j \equiv - e \frac{d {\cal N}_j}{dt}$, and ${\cal N}_j$
  is the number of electrons in contact $j$.}
\bibitem[{Rem({\natexlab{b}})}]{Remark05a}
\bibinfo{note}{In particular the first term in Eq.~(\ref{eq:jom}) is $\Omega$
  independent, and vanishes after differentiation. Notice that when the term
  $\left\langle \frac{\partial \hat{J}_j}{\partial X} \right\rangle$ does not
  vanish, a counter term from the comutator in Eq.~(\ref{eq:jom}) cancels it so
  that at $\Omega =0$ the total current is zero.}
\bibitem[{Rem({\natexlab{c}})}]{Remark05f}
\bibinfo{note}{The results of Ref.~\cite{Sharmab03}, e.g., Eq.~(23) follow as a
  special case. Using their notation, in that case $\frac{d H}{ d t}=\delta
  \dot{\phi} J_{\Gamma_0}$.}
\bibitem[{Rem({\natexlab{d}})}]{Remark05c}
\bibinfo{note}{Here C is the Keldysh contour, $\eta(\eta')=\pm$ are the Keldysh
  branches and the times $t_1(t_1')$ belong to branch $\eta(\eta')$
  respectively. The retarted (r), advanced (a) and lesser ($<$) Green functions
  are related to the Keldysh functions via: $G^<=G^{-+}$, $G^r=G^{--} -G^{<}$,
  $G^a=G^{<} -G^{++}$. In the subscripts of the Green function we use $c_{k
  \alpha}(d_n) \rightarrow k \alpha(n)$.}
\bibitem[{Rem({\natexlab{e}})}]{Remark05d}
\bibinfo{note}{Under the assumption $V_{k \alpha,n}(X)=V_{ \alpha,n}
  \lambda_\alpha(X)$, the first term in Eq.~(\ref{eq:eml}) vansihes
  \cite{Hewson93d} and the second part with the square brackets is equal to the
  RHS of Eq.~(8) of Ref.~\cite{Fazio05short} (divided by 2 due to spin).
  However the terms containing $\mathbf{\Lambda}_X(\omega,\Omega)$ are
  neglected in the average time approximation of Ref.~\cite{Fazio05short}.}
\bibitem[{Rem({\natexlab{f}})}]{Remark05k}
\bibinfo{note}{Since at $T=0$ the system is a FL except at $(h,\Delta)=(0,0)$,
  this result can also be derived from Brouwer formula~\cite{Brouwer98}, where
  the scattering matrix is derived using phase shifts extracted for example
  from NRG studies~\cite{Pustilnik04short}}.
\end{thebibliography}

\end{document}